\documentstyle[manuscript,eqsecnum,aps]{revtex}
\begin{document}
\def\cf{{\it cf.\ }}
\def\eff{{\rm eff}}
\def\ed{{\sl ed.~by\ }}
\def\eg{{\it e.g.,\ }}
\def\Eg{{\it E.g.,\ }}
\def\ie{{\it i.e.,\ }}
\def\Ie{{\it I.e.,\ }}
\def\etal{{\it et al.~}}
\def\etc{{\it etc.}}
\def\via{{\it via}}
\def\Eq#1{Eq.~(#1)}                     %       ditto
\def\eq#1{Eq.~(#1)}                     %       ditto
\def\Eqs#1{Eqs.~(#1)}                   %       ditto
\def\eqs#1{Eqs.~(#1)}                   %       ditto
\def\frac#1#2{{\textstyle #1 \over \textstyle #2}}
\def\half{{\textstyle {1 \over 2}}}
\gdef\journal#1, #2, #3, 1#4#5#6{               % Journal reference.
    {\sl #1~}{\bf #2}, #3, (1#4#5#6)}           % Comma sets off:
						% name, vol, page,
						% (year).
\draft
\date{\today}

\title{Unitarity and Causality in Generalized
 Quantum Mechanics for Non-Chronal Spacetimes}

\author{James B.
Hartle\thanks{hartle@cosmic.physics.ucsb.edu}}
\vskip .13 in
\address{Department of Physics, University of California,
Santa Barbara, CA
93106-9530}

\maketitle

\begin{abstract}

Spacetime must be foliable by spacelike surfaces for the quantum
mechanics of matter fields to be formulated in terms of
a unitarily evolving state vector defined on spacelike surfaces.  When a
spacetime cannot be foliated by spacelike surfaces, as in the case of
spacetimes with closed timelike curves, a more general formulation of
quantum mechanics is required.  In such generalizations the
transition matrix between alternatives in regions of spacetime
 where states {\it can} be defined may
be non-unitary.  This paper describes a generalized quantum mechanics
whose probabilities consistently obey the rules of probability
theory even in the presence of such non-unitarity.  The usual notion of
state on a spacelike surface is lost in this generalization and
familiar notions of causality are modified.  There is no signaling
outside the light cone, no non-conservation of energy,
no ``Everett phones'',
 and probabilities of present events do not depend
on particular alternatives of the future.  However, the generalization
is acausal
in the sense that the existence of non-chronal regions of
spacetime in the future can affect the probabilities of alternatives
today. The detectability of non-unitary evolution and violations of
causality in measurement situations are briefly considered.  The
evolution of information in non-chronal spacetimes is described.
\end{abstract}

\pacs{}

\setcounter{footnote}{0}
\section{Introduction}
\label{sec:intro}

Conventional formulations of the quantum mechanics of matter fields
in a curved background spacetime require that this spacetime be foliable
by a family of spacelike surfaces.  A family of spacelike surfaces
is needed just to define a state of the matter fields on a
spacelike surface and the progress of this state into the future by
either unitary evolution between spacelike surfaces or by ``state vector
reduction'' on them.  However, not all spacetimes admit a foliation by
spacelike surfaces.  For example, spacetimes with closed timelike curves,
such as would be produced by the motion of wormhole mouths,
permit no foliating family of spacelike surfaces \cite{WORM}.
The quantum mechanics of matter
fields in spacetimes with such non-chronal regions therefore cannot be
formulated in terms of the evolution of
states on  spacelike surfaces.
 Rather, a more general formulation of quantum mechanics is
required.  Generalizations based on the ideas of quantum computation
have been described by Deutsch \cite{Deu91} and generalizations based on
the algebraic approach to field theory have been discussed by Yurtsever
\cite{Yur93}. Here, we pursue another class of generalizations based on
the sum-over-histories formulation of quantum theory.
 Generalizations of this kind have previously
been discussed by Klinkhammer and Thorne \cite{KTpp}, Friedman, Simon,
and Papastamatiou \cite{FPS91a} and the author \cite{Har91a}.
Specifically, we explore the notions of unitarity and causality and the
connections between them in this class of generalizations.

Feynman's sum-over-histories formulation of quantum mechanics is a
natural route to a generalized quantum mechanics of matter fields in
spacetimes with non-chronal regions because, with it, quantum mechanics may
be cast into a fully spacetime form that does not employ a notion of
state on a foliating family of spacelike surfaces \cite{KTpp,FPS91a,Har91a}.
For example, in  the sum-over-histories formulation,
 quantum dynamics is expressed, not through a
differential equation, but rather by giving
the amplitude for
a fine-grained field history --- a four-dimensional field
configuration, $\phi(x)$. In Feynman's prescription this amplitude
 is proportional
to
\begin{equation}
\exp\bigl(iS[\phi(x)]/\hbar\bigr)\label{oneone}
\end{equation}
where $S$ is the action functional for the field.  Quantum dynamics
can be defined in this way
even when spacetime contains
non-chronal regions.  The alternatives potentially
assigned probabilities by quantum theory can also be described
four-dimensionally as partitions, or
coarse grainings, of the fine-grained field histories into classes.  For
instance, the four-dimensional field histories could be partitioned by
the values of the field configurations $\phi ({\bf x})$ on a
spacelike surface $\sigma$.  The amplitudes for such alternatives define
state functionals on $\sigma$ in familiar quantum theories formulated in
terms of states on spacelike surfaces.
However, even in non-chronal
regions, where there are no foliating families of spacelike surfaces, we
can still define meaningful coarse-grainings of four-dimensional field
configurations.  For example, we could partition the field histories by
the value of a field averaged over a region of spacetime deep inside a
wormhole throat.  A decoherence functional defining the interference
between such individual alternatives may be defined and the probabilities for
decohering sets of alternatives calculated.  In this way the quantum
theory of fields may be put into fully four-dimensional form
free from the need of
a foliating family of spacelike surfaces
\cite{Har91a,Har91b,Har93c,Harpp}.

If the non-chronal regions of spacetime are bounded, then the spacetime
contains initial and final regions before and after the non-chronal one
 in which
familiar alternatives of the spatial field configurations can be
defined on spacelike surfaces
(Figure 1).
Transition probabilities between such alternatives are of interest.
Transition amplitudes between a definite
spatial field configuration, $\phi^\prime({\bf x})$, on an initial
spacelike surface $\sigma^\prime$ and a configuration $\phi^{\prime\prime}
({\bf x})$ on a
final surface $\sigma^{\prime\prime}$ are given by a
sum-over-histories expression of the form
\begin{equation}
\bigl\langle\phi^{\prime\prime}({\bf x}), \sigma^{\prime\prime} |
\phi^\prime ({\bf x}), \sigma^\prime\bigr\rangle =
\int\nolimits_{[\phi^\prime, \phi^{\prime\prime}]}\delta\phi
\ \exp\bigl(iS[\phi(x)]/\hbar\bigr)\ .\label{onetwo}
\end{equation}
The sum is over four-dimensional field configurations between
$\sigma^\prime$ and $\sigma^{\prime\prime}$ that match the prescribed
spatial configurations on those surfaces.  By such methods, for example,
an $S$-matrix for scattering through spacetime regions with closed
timelike curves could be defined and calculated.

When spacetime can be foliated by a family of
 spacelike surfaces, eq.~(\ref{onetwo})
coincides with the unitary evolution operator generated by the
Hamiltonian for the family.   That is because, as Dirac \cite{Dir33}
 and Feynman \cite{Fey48} showed, when two
spacelike surfaces are close together, the matrix elements of the
operator effecting unitary evolution between them is proportional to
$\exp(iS)$ where $S$ is the action of the classical field history
interpolating between the two.  Explicitly in the case of two
constant-time surfaces in Minkowksi space

\begin{equation}
\exp\left\{iS\left[\phi^{\prime\prime}({\bf x}), t^{\prime\prime};
\phi^\prime({\bf x}), t^\prime\right]/\hbar\right\}
\propto\left\langle\phi^{\prime\prime} ({\bf x}) | \exp \left[-i
H(t^{\prime\prime}-t^\prime)/\hbar\right] | \phi^\prime({\bf
x})\right\rangle\ .\label{onethree}
\end{equation}
However, in the absence of a connection like (\ref{onethree}),
or even a well
defined meaning for its right hand side, there is no particular reason
to expect a construction like (\ref{onetwo}) to yield a unitary transition
matrix.\footnote{A conclusion also reached by Deutsch \cite{Deu91} from the
point of view of quantum computation.}  Calculations by Klinkhammer and
Thorne \cite{KTpp} in non-relativistic quantum mechanics first suggested
that the evolution defined by (\ref{onetwo}) might be non-unitary.
General results of Friedman,
Papastamatiou and
Simon \cite{FPS91a,FPS91b} in field theory, and  explicit examples of
Boulware \cite{Bou92} and Politzer
\cite{Pol92}, show the following: The scattering matrix
constructed from the sum-over-histories (\ref{onetwo}) is unitary for
free field theories in spacetimes with
closed timelike curves, but not, in
general for interacting theories,
 order by order in perturbation theory.\footnote{Goldwirth, Perry and
Piran \cite{GPP93} concluded that even free theories were non-unitary.
This was corrected in \cite{GPPT93}, where some of the results of
Klinkhammer and Thorne \cite{KTpp} for free theories are included.}
 This paper discusses the
implications of this non-unitarity.

Even were spacetime foliable by spacelike surfaces it would still be
difficult to reconcile non-unitary evolution with the notion of state on
a spacelike surface.  The reasons, stated clearly by T. Jacobson \cite{Jac91},
 are reviewed in Section II.  However, a generalized quantum mechanics
neither requires, nor does it always permit, a notion of ``state on a
spacelike surface''.  In Section III we spell out enough details of the
generalized sum-over-histories quantum mechanics sketched above to show
how it consistently incorporates non-unitary evolution represented by
the transition matrix (\ref{onetwo}) without employing a notion of
state. The price for this generalization
is not only the absence of a notion of state on a spacelike surface, but
also a violation of causality that is discussed
in Section IV.  There is no violation of causality in the sense that
signals propagate outside the light cone.  Neither do probabilities in
the present depend on specific alternatives in the future.  However, the
existence of future non-chronal regions of spacetime will influence
probabilities in the present.  A theory of the future geometry of
spacetime, as well as of the
initial condition of the closed system and the geometry up to the
present,
is thus required for present prediction.

The theory of laboratory scattering measurements in the presence of
non-chronal regions is developed in Section
V and used  to give a preliminary discussion of
how violations of unitarity and causality might
be detected.  In Section VI we introduce a notion of spacetime
information that is not tied to a notion of state on a spacelike surface
and shows how it evolves in a spacetime with non-chronal regions. Section
VII shows that various anomalies such as non-conservation of energy,
signaling faster than light, and communication between non-interfering
branches that exist in some other generalizations of quantum mechanics
are absent
from this one.

\section{Non-Unitarity and the Quantum Mechanics of States}
\label{sec:uni}

In its simplest interpretations, non-unitary evolution of a quantum
state defined on spacelike surfaces is either inconsistent, or, as shown
by T. Jacobson \cite{Jac91}, dependent on the choice of spacelike
surfaces.  This section briefly reviews these arguments.

We consider a fixed background spacetime containing a bounded,
non-chronal
region $NC$, as shown in Figure 1.  Consider an initial state $|\psi
(\sigma^\prime)\rangle$ on a spacelike surface $\sigma^\prime$
before\footnote{We shall be more precise about the meanings of
``before'' and ``after'' in Section III.}
$NC$.  Suppose its evolution to a spacelike surface
$\sigma^{\prime\prime}$ after $NC$ is given by a
non-unitary evolution operator $X$:
\begin{equation}
|\psi (\sigma^{\prime\prime})\rangle = X | \psi (\sigma^\prime)
\rangle\ . \label{twoone}
\end{equation}
We now consider the calculation of the probabilities of an exhaustive
set of exclusive alternatives on this spacelike surface represented
by a set of (Schr\"odinger-picture) projection operators satisfying
\begin{equation}
\sum\nolimits_\alpha P_\alpha = 1\quad , \quad P_\alpha P_\beta =
\delta_{\alpha\beta} P_\beta\ . \label{twotwo}
\end{equation}
What rule should be used to calculate these probabilities?

The usual prescription for the probability of the alternative
corresponding to $P_\alpha$ on $\sigma$ is
\begin{equation}
p(\alpha; \sigma) = \parallel P_\alpha | \psi (\sigma) \rangle
\parallel^2\ . \label{twothree}
\end{equation}
If $|\psi(\sigma^\prime)\rangle$
is normalized
so that
\begin{equation}
\sum\nolimits_\alpha p(\alpha; \sigma^\prime) = 1\ , \label{twofour}
\end{equation}
then (\ref{twoone}) will imply for the probabilities of the same
alternatives on the later surface
\begin{equation}
\sum\nolimits_\alpha p(\alpha; \sigma^{\prime\prime}) = \left\langle
\psi(\sigma^\prime) | X^\dagger X | \psi (\sigma) \right \rangle \not= 1
\ .
\label{twofive}
\end{equation}
When $X$ is not unitary, probability is not conserved and
the prescription (\ref{twothree}) for
assigning probabilities is thus inconsistent.

The generalization of (\ref{twothree})
\begin{equation}
p(\alpha; \sigma) = \frac{\parallel P_\alpha | \psi (\sigma)\rangle
\parallel^2}{\parallel | \psi (\sigma) \rangle \parallel^2}
\label{twosix}
\end{equation}
suggests itself as a way of maintaining the
requirement that the probabilities of an exhaustive set of alternatives
sum to unity.  However, Jacobson \cite{Jac91} has shown that this rule is
not covariant with respect to the choice of spacelike surfaces.  Consider
a set of alternatives $\{P_\alpha(R)\}$ that distinguish only properties
of fields on a spacelike surface that are
restricted to a region $R$ that is spacelike
separated from a non-chronal region $NC$.
For example, an exhaustive set of ranges of
the average of a field
over $R$ defines one such set of alternatives.
Since $R$ is spacelike to $NC$ it may be considered either as a part of
 a spacelike surface
$\sigma^\prime$ that is before $NC$ or as part of a spacelike surface
$\sigma^{\prime\prime}$ that is after $NC$ (Figure 2).
According to (\ref{twosix}) and
(\ref{twoone}), the probabilities for the alternatives
$\{\alpha\}$ evaluated on
$\sigma^\prime$ would be
\begin{equation}
p(\alpha; \sigma^\prime) = \frac{\left\langle \psi (\sigma^\prime) |
P_\alpha (R) | \psi (\sigma^\prime)\right\rangle}{\left\langle
\psi(\sigma^\prime)  | \psi(\sigma^\prime)\right\rangle}\ ,
\label{twoseven}
\end{equation}
while on $\sigma^{\prime\prime}$ they would be
\begin{equation}
p(\alpha; \sigma^{\prime\prime}) = \frac{\left\langle \psi
(\sigma^\prime)| X^\dagger P_\alpha (R) X | \psi
(\sigma^\prime)\right\rangle}{\left\langle\psi(\sigma^\prime) |
X^\dagger X | \psi (\sigma^\prime)\right\rangle}\ .
\label{twoeight}
\end{equation}
These must be equal since the alternatives are the same.

A state on $\sigma^\prime$ that is an eigenvector of the
field configuration $\phi
({\bf x})$ with ${\bf x}\epsilon R$,  evolves into a state on
$\sigma^{\prime\prime}$ that is also an eigenvector of $\phi ({\bf x})$,
with ${\bf x} \epsilon R$, having the same eigenvalue.  Thus,
\begin{equation}
\bigl[X, P_\alpha(R)\bigr] = 0\ .
\label{twonine}
\end{equation}
Were $X$ unitary, (\ref{twonine}) would imply the equality of the
numerators in (\ref{twoseven}) and (\ref{twoeight}) and of the
denominators.  However, when $X$ is non-unitary the expressions
(\ref{twoseven}) and (\ref{twoeight}) cannot be equal for all
states $|\psi(\sigma)\rangle$.
Non-unitary evolution
therefore implies that the probabilities for the alternatives
$\{P_\alpha(R)\}$ are different on $\sigma^\prime$ and
$\sigma^{\prime\prime}$.  Quantum mechanics defined by the rule
(\ref{twosix}) is not covariant with respect to the choice of spacelike
surfaces unless the evolution is unitary.  This is the essence of
Jacobson's argument.

Thus a non-unitary transition matrix, say constructed by a
sum-over-histories as in (\ref{onetwo}), cannot be used to construct a
quantum mechanics in which probabilities are computed from a ``state of
the system on a spacelike surface'' using either of the prescriptions
(\ref{twoone}) or (\ref{twosix}) if we insist on covariance with
respect to the choice of spacelike surfaces.
In the next Section we shall show how a
generalized quantum mechanics that avoids this problem
can be constructed incorporating
such non-unitary
transition matrices.  Such generalizations will not, of course, admit a
notion of ``state on a spacelike surface'' in any of the senses
discussed in this Section.

\section{Generalized Quantum Mechanics}
\label{sec:gen}

In this Section we will spell out more concretely some details of
generalized quantum theories that consistently incorporate non-unitary
evolution.  We have described the principles of generalized quantum
mechanics elsewhere \cite{Har91a,Harpp} and do not review them here.
The sum-over-histories quantum mechanics for non-chronal spacetimes that
was sketched in the Introduction is one example of a generalized quantum
mechanics incorporating non-unitary evolution.  However, it turns out
that with respect to alternatives defined on spacelike surfaces where
they exist, and the
transitions between them, a more general discussion can be given which
is largely independent of the specific mechanism of non-unitarity [for
example that of (\ref{onetwo})].  We shall exhibit this general framework.

We are concerned most generally with the quantum mechanics of a closed
system containing both observers and observed, measuring apparatus and
measured subsystems. In the present investigation, the  closed system is
an interacting quantum field theory in a {\it fixed} given,
 background spacetime
geometry.  To keep the notation manageable we shall
consider a single, scalar field, $\phi(x)$.  We shall assume
that spacetime outside a bounded region $NC$ is foliable by spacelike
surfaces (see Figure 3).  Thus we can identify an initial region
${\cal IN}(NC)$ outside of $NC$, no point of which can be reached from
any point of $NC$ by a timelike curve that is future pointing outside of
$NC$.  ${\cal IN}(NC)$ is foliable by spacelike surfaces.
Similarly we can define a final region
${\cal FN}(NC)$ [generally overlapping ${\cal IN}(NC)$]
that is foliable by spacelike surfaces.
We will loosely refer to ${\cal IN}(NC)$ as ``before''
$NC$ and ${\cal FN}(NC)$ as ``after'' $NC$. Later we can consider the
case of several disjoint non-chronal regions.

The most general objective of a quantum theory is the prediction of the
probabilities of the individual histories in an exhaustive set of
alternative, coarse-grained histories of the closed system.  As mentioned
above,  we shall restrict attention in this paper
to sets of histories consisting of alternatives defined  on
spacelike surfaces foliating the initial and final regions.  This has
the advantage that the usual apparatus of operators on Hilbert space may
be used to describe these alternatives.\footnote{For discussion of
more general classes of spacetime
alternatives see, e.g. \cite{Har91a,Har91b,YTxx}}

In the Schr\"odinger picture, an exhaustive and exclusive set of
alternatives defined on a spacelike surface corresponds to a set of
projection operators $\{P_\alpha\}$ satisfying (\ref{twotwo}).
The $P_\alpha$, for example, might be projections onto ranges of values
a field averaged over a spatial region $R$ in the surface.  Specifying
(generally different) sets of alternatives $\{P^1_{\alpha_1}\},
\{P^2_{\alpha_2}\}, \cdots, \{P^n_{\alpha_n}\}$ on a sequence of
non-intersecting
spacelike
surfaces $\sigma_1, \cdots, \sigma_n$ defines a set of coarse-grained
alternative histories for the system.  A particular history corresponds
to a particular sequence of alternatives $\alpha_1, \cdots, \alpha_n$,
that we shall often abreviate by a single index, {\it viz}: $\alpha =
(\alpha_1,
\cdots, \alpha_n)$.  The exhaustive set of histories
 consists of all possible sequences $\{\alpha\}$.
The histories are coarse grained because not all information is
specified that
could be specified.  Alternatives are not specified at
each and every time, and the alternatives that are specified do not
correspond to a complete set of states unless all the $P$'s are
one-dimensional.

A quantum theory of a closed system does not assign probabilities to
every set of coarse-grained histories of a closed system.  In the
two-slit experiment, for example, we cannot assign probabilities to the
alternative histories in which the electron went through one slit or the
other and arrived at a definite point on the detecting screen.  It would
be inconsistent to do so because, as a consequence of quantum mechanical
interference, these probabilities would
not correctly sum to the probability to arrive at the designated point
on the screen.  The quantum mechanics of closed systems assigns
probabilities only
to the members of sets of alternative, coarse-grained histories for which there
is negligible interference between the individual histories in the set
as a consequence of the system's
dynamics and boundary conditions \cite{Gri84,Omnsum,GH90a}.
Such sets of histories are said to {\it decohere}.  In a generalized quantum
theory, the interference between histories in a set
is measured by a decoherence functional
incorporating information about the system's dynamics and initial
condition. The decoherence functional, $D(\alpha^\prime, \alpha)$, is a
complex function of pairs of histories
satisfying certain general conditions that  we shall describe below. The set
decoheres if $D(\alpha^\prime, \alpha)$ is
sufficiently small for {\it all} pairs of {\it different} histories in
the set $\{\alpha\}$.
When that is the case, the probabilities of the individual
histories $p(\alpha)$ are the diagonal elements of $D(\alpha^\prime,
\alpha)$.  The rule both for when probabilities may  be assigned
to a set of coarse-grained histories and what these probabilities are
may thus be summarized by the fundamental formula:
\begin{equation}
D(\alpha^\prime, \alpha) \approx \delta_{\alpha^\prime\alpha} p(\alpha)
\ . \label{threeone}
\end{equation}

When spacetime is completely foliable by spacelike surfaces, the
decoherence functional of familiar Hamiltonian quantum mechanics is
given by
\begin{eqnarray}
D(\alpha^\prime, \alpha)&=& Tr\Bigl[P^n_{\alpha^\prime_n} U(\sigma_n,
\sigma_{n-1}) P^{n-1}_{\alpha^\prime_{n-1}} U(\sigma_{n-1}, \sigma_{n-2})
\nonumber\\
\cdots P^1_{\alpha^\prime_1} U(\sigma_1, \sigma_0) \rho& U&(\sigma_0,
\sigma_1) P^1_{\alpha_1} \cdots
U(\sigma_{n-2}, \sigma_{n-1}) P^{n-1}_{\alpha_{n-1}} U(\sigma_{n-1},
\sigma_n) P^n_{\alpha_n}\Bigr] \label{threetwo}
\end{eqnarray}
where $\rho$ is the density matrix describing the initial condition of
the system of fields on an initial spacelike surface, $\sigma_0$,
and $U(\sigma^{\prime\prime}, \sigma^\prime)$ is the
unitary evolution operator between spacelike surfaces $\sigma^\prime$ and
$\sigma^{\prime\prime}$.

Generalizing the form of the decoherence functional (\ref{threetwo})
generalizes Hamiltonian quantum mechanics.  A wide class of
generalizations called generalized quantum theories \cite{Har91a,Harpp}
 have decoherence
functionals that ~(i) are Hermitian: $D(\alpha, \alpha^\prime) =
D^*(\alpha^\prime, \alpha)$, (ii) are normalized:
$\Sigma_{\alpha\alpha^\prime} D(\alpha, \alpha^\prime)=1$, (iii) have
positive diagonal elements: $D(\alpha, \alpha) \geq 0$, and, most
importantly, (iv) obey the principle of superposition in the following sense:
A coarse graining of the set
$\{\alpha\}$ means a partition of that set into a new set of (generally
larger) exhaustive and
exclusive classes, $\{\bar\alpha\}$. A decoherence functional
satisfies the principle of superposition when
\begin{equation}
D(\bar\alpha^\prime, \bar\alpha) =
\sum_{\alpha^\prime\epsilon\bar\alpha^\prime}
\sum_{\alpha\epsilon\bar\alpha} D(\alpha^\prime, \alpha)
\label{threethree}
\end{equation}
for all coarse grainings $\{\bar\alpha\}$ of $\{\alpha\}$.  When a set
of histories decoheres, and probabilities are assigned according to the
fundamental formula (\ref{threeone}), the numbers $p(\alpha)$ lie between 0
and 1 and satisfy the most general form of the probability sum rules
\begin{equation}
p(\bar\alpha) = \sum_{\alpha\epsilon\bar\alpha} p(\alpha)\ .
\label{threefour}
\end{equation}
The decoherence functional of Hamiltonian quantum mechanics,
(\ref{threetwo}), is easily seen to satisfy ~(i) --- (iii), and satisfies
{\it sums} of the projections in the fine-grained set.

Suppose we consider a spacetime and a single non-chronal region $NC$ and
restrict attention to alternatives defined on spacelike surfaces either
entirely in the region ${\cal IN}(NC)$ ``before'' $NC$ or in the region
${\cal FN}(NC)$ ``after'' it.  Suppose the evolution between a spacelike
surface $\sigma_-$ before $NC$ and a spacelike surface $\sigma_+$ after
$NC$ is
not described by a unitary matrix $U$, but by a non-unitary matrix $X_S$.
The decoherence functional (\ref{threetwo}) with $U$ replaced by $X_S$
no
longer satisfies the general requirements ~(i) --- (iv).  However, the
following generalization does satisfy them:
\begin{eqnarray}
&D(\alpha^\prime, \alpha) = N\ Tr \Bigl[P^n_{\alpha^\prime_n}
U(\sigma_n,
\sigma_{n-1}) \cdots P^{k+1}_{\alpha^\prime_{k+1}} U(\sigma_{k+1},
\sigma_+) X_S
U (\sigma_-, \sigma_k) P^k_{\alpha^\prime_k} \cdots
U(\sigma_2, \sigma_1)
P^1_{\alpha^\prime_1}
\nonumber\\
& U(\sigma_1, \sigma_0)\ \rho
\, U(\sigma_0, \sigma_1)
 P^1_{\alpha_1} U(\sigma_1, \sigma_2)
 \cdots P^k_{\alpha_k} U (\sigma_k, \sigma_-)
X^\dagger_S U(\sigma_+, \sigma_{k+1}) P^{k+1}_{\alpha_{k+1}} \cdots
U(\sigma_{n-1}, \sigma_n) P^n_{\alpha_n}\Bigr] \label{threefive}
\end{eqnarray}
where
\begin{equation}
N^{-1} = Tr \left(X \rho X^\dagger \right) \label{threesix}
\end{equation}
and $\sigma_1, \cdots, \sigma_k$ lie before $\sigma_-$ in ${\cal
IN}(NC)$
while
$\sigma_{k+1}, \cdots \sigma_n$ lie after $\sigma_+$ in ${\cal FN}(NC)$.

The expression (\ref{threefive}) may be simplified by introducing a kind
of
Heisenberg picture with operators
\begin{mathletters}
\label{allequations}
\begin{equation}
P^i_{\alpha_i} (\sigma_i) = U^{-1} (\sigma_i, \sigma_0) P^i_{\alpha_i}
U(\sigma_i, \sigma_0)\quad , \quad \sigma < \sigma_- \label{threesevena}
\end{equation}
\begin{equation}
P^i_{\alpha_i}(\sigma_i) = U^{-1}(\sigma_i, \sigma_f) P^i_{\alpha_i}
U(\sigma_i, \sigma_f)\quad , \quad \sigma > \sigma_+ \label{threesevenb}
\end{equation}
\end{mathletters}
and
\begin{equation}
X = U^{-1}(\sigma_+, \sigma_f) X_S U(\sigma_-, \sigma_0)\ .
\label{threeeight}
\end{equation}
where $\sigma_f$ is a final surface in the far future.  Then
(\ref{threefive})
is
\begin{eqnarray}
D(\alpha^\prime, \alpha)& = &N\ Tr\Bigl[P^n_{\alpha^\prime_n} (\sigma_n)
\cdots P^{k+1}_{\alpha^\prime_{k+1}} (\sigma_{k+1}) X
\ P^k_{\alpha^\prime_k} (\sigma_k)
\nonumber\\
\cdots P^1_{\alpha^\prime_1} (\sigma_1)& \rho& P^1_{\alpha_1}(\sigma_1)
\cdots P^k_{\alpha_k}(\sigma_k)
X^\dagger
P^{k+1}_{\alpha_{k+1}} (\sigma_{k+1}) \cdots P^n_{\alpha_n} (\sigma_n)
\Bigr]\ .
\label{threenine}
\end{eqnarray}
The expression can be written
even more compactly if we introduce the notation
\begin{equation}
C_\alpha = P^k_{\alpha_k}(\sigma_k) \cdots P^1_{\alpha_1}(\sigma_1)
\label{threeten}
\end{equation}
for a chain of projections on spacelike surfaces before $\sigma_-$,
and
\begin{equation}
C_\beta = P^n_{\beta_n}(\sigma_n) \cdots P^{k+1}_{\beta_{k+1}}
(\sigma_{k+1})
\label{threeeleven}
\end{equation}
for a chain on spacelike surfaces after $\sigma_+$. Then
\begin{equation}
D\left(\beta^\prime, \alpha^\prime ; \beta, \alpha\right) = \frac{Tr
\left(C_{\beta^\prime} X C_{\alpha^\prime} \rho C^\dagger_\alpha
X^\dagger C^\dagger_\beta\right)}{Tr \left(X\rho X^\dagger\right)} \ .
\label{threetwelve}
\end{equation}

The decoherence functional (\ref{threetwelve}) defines a quantum
mechanics
that reduces to the usual one (\ref{threetwo}) when the evolution is
unitary,
but generalizes it when it
is not.  It consistently assigns
probabilities to decoherent sets of histories.  There is no issue of the
violation of a probability sum rule like (\ref{twofive}) here.  All
probability sum rules (\ref{threefour}) are satisfied as a consequence of
decoherence including the elementary requirement that the probabilities
of an exhaustive set of alternatives sum to $1$.
Neither is there hypersurface dependence of local
probabilities as with Jacobson's rule (\ref{twosix}).
 From (\ref{threetwelve}) it
follows that the probability of a set of alternatives $P_\alpha(R)$ that
distinguish only field values on a local piece of a spacelike surface
$R$ that is everywhere spacelike separated from the
non-chronal region $NC$ are
\begin{equation}
p(\alpha, \sigma^\prime) = N\ Tr\left[XP_\alpha (R) \rho P_\alpha (R)
X^\dagger\right] \label{threethirteen}
\end{equation}
when $R$ is considered part of a spacelike surface $\sigma^\prime$ to the
before $NC$, and given by
\begin{equation}
p(\alpha; \sigma^{\prime\prime}) = N\ Tr\left[P_\alpha(R) X \rho
\, X^\dagger P_\alpha (R)\right] \label{threefourteen}
\end{equation}
when $R$ is considered part of a spacelike surface after $NC$.
However, since $P_\alpha(R)$ and $X$ commute [\cf Eq.~(\ref{twoeight})], eqs
(\ref{threethirteen}) and (\ref{threefourteen}) are equivalent.
The rule (\ref{twosix}) includes or does not include
the non-unitary evolution operator $X$ depending on which surface is
chosen.  By contrast the rules (\ref{threethirteen}) and
(\ref{threefourteen}) both include an $X$.  The order of the $X$ with
respect to projection operator representing the alternative in $R$ is
different depending on whether $R$ is considered a part of
$\sigma^\prime$ or $\sigma^{\prime\prime}$, but that order is immaterial
since the operators commute. The
generalized quantum mechanics defined by the decoherence functional
(\ref{threetwelve}) is thus consistent with elementary requirements.  In
the following we shall explore its consequences.

\section{Causality}
\label{sec:caus}

The past influences the future but the future does not influence the
past; that is the essence of causality.  A fixed spacetime geometry
whose causal structure defines ``future'' and ``past'' as needed just to
ask whether a theory is consistent with causality or
not.  A fixed background spacetime {\it has} been assumed for the field
theories that are the concern of this paper,  but the future and past cannot
be unambiguously distinguished for points {\it inside} non-chronal regions
connected by closed timelike curves.  However, we {\it can} ask
whether the probabilities of a set of alternatives defined entirely
{\it outside} such regions are independent of the geometry of spacetime to
their future.  It is straightforward to see that the generalized quantum
mechanics of matter fields described in the previous Section is {\it
not} causal in this sense if the evolution through non-chronal
regions is
not unitary.

Suppose that spacetime contains a single non-chronal region that is to our
future and we are concerned with the probabilities of a chain of
alternatives $C_\alpha$ all occurring {\it before} the non-chronal region.
If these alternatives decohere, then their probabilities $p(\alpha)$ are
given, according to (\ref{threeone}) and (\ref{threetwelve}) by
\begin{equation}
p(\alpha) = N\ Tr \bigl(XC_\alpha\rho C^\dagger_\alpha X^\dagger\bigr)
\label{fourone}
\end{equation}
where $X$ describes the evolution through the non-chronal region and $N^{-1}
= Tr(X\rho X^\dagger)$.  Were $X$ unitary, the cyclic property of the
trace could then be used to show
\begin{equation}
p(\alpha) = Tr\bigl(C_\alpha\rho C^\dagger_\alpha\bigr)\ . \label{fourtwo}
\end{equation}
Eq.~(\ref{fourtwo}) could then be written out in the Schr\"odinger picture
using
(\ref{threefive}).  Since only $U(\sigma,
\sigma_0)$ for values of $\sigma$ less than the last $\sigma_n$ occur
in the
chain $C_\alpha$, there is no dependence on the geometry of spacetime to
the future of the surface $\sigma_n$, whether or not it contains
non-chronal regions.
  In this sense, unitary evolution leads to causality.

If $X$ is not unitary then the probabilities defined by
(\ref{fourone}) depend on the future geometry of spacetime.  Experiments
could, in principle, detect the existence of non-chronal regions
in our future by testing whether present data is better fit by
(\ref{fourtwo}) or (\ref{fourone}) with the appropriate $X$.
We shall return to  some simple considerations of such experiments in
Section V.

Another way of seeing that information about the future is required to
calculate present probabilities is to write $\rho_f=X^\dagger X$ and use
the cyclic property of the trace to reorganize (\ref{fourone}) as
\begin{equation}
p(\alpha)= N\ Tr\bigl(\rho_f C_\alpha \rho C^\dagger_\alpha\bigr)
\label{fourthree}
\end{equation}
where now $N^{-1} = Tr(\rho_f\rho)$. Eq.~(\ref{fourthree}) is the formula for
the probabilities of a generalized quantum mechanics with both an
initial condition $\rho$ and a final condition $\rho_f$.  Such
generalizations were discussed in \cite{Gri84} and \cite{GH93b}
 for the quantum mechanics
of closed systems.  Information about {\it both}
 the future
and the past is required to make predictions in the present.  In the
example under discussion, that information concerns the failure of
unitarity in the future arising from non-chronal regions of spacetime.

The notion of state of the system on a spacelike surface provides the
most familiar expression of causality in usual quantum mechanics.  From a
knowledge of the state in the present, all future probabilities may be
predicted.  Thus the present determines the future. We next show that
the generalized quantum mechanics under discussion does not contain such
a notion of state.

When the probability formula is the usual (\ref{fourtwo}), it is
straightforward reformulate it in terms of states on spacelike surfaces.
Let $\sigma$ denote the
spacelike surface defining the present, let $C_\alpha$ denote a history
of alternatives that have already happened, and $C_\beta$ a history of
future alternatives whose probabilities we wish to predict.  The
conditional probability for the future alternatives given the past ones
is,
\begin{equation}
p(\beta|\alpha) = p(\beta, \alpha)/p(\alpha)\ .
\label{fourfour}
\end{equation}
If the joint probabilities on the right hand side
of (\ref{fourfour}) are given by (\ref{fourtwo}), then $p(\beta|\alpha)$ can be
written in terms of an effective density matrix $\rho_{\rm eff}$
defined on
$\sigma$ as
\begin{equation}
p(\beta|\alpha) = Tr\bigl[C_\beta \rho_{\rm eff} (\sigma)
C^\dagger_\beta\bigr]\ ,
\label{fourfive}
\end{equation}
where
\begin{equation}
\rho_{\rm eff} (\sigma) =
\frac{C_\alpha\rho C^\dagger_\alpha}{Tr\bigl(C_\alpha\rho
C^\dagger_\alpha\bigr)}\ . \label{foursix}
\end{equation}

The density matrix $\rho_{\rm eff} (\sigma)$ is the usual notion of
state on a spacelike surface.  As $\sigma$ advances, $\rho_{\rm eff}
(\sigma)$ is constant in
time in this Heisenberg picture until the time of a new alternative is
reached at which point it is ``reduced'' by the addition of a new
projection to the chain $C_\alpha$.  The conditional probabilities of
future decoherent alternatives continue to be given by
(\ref{fourfive}) with the new $\rho_{\rm eff} (\sigma)$.

If the probability formula is (\ref{fourone}) or (\ref{fourthree}) then it is
{\it not}
possible to construct a $\rho_{\rm eff}$ on a spacelike surface from
which alone future probabilities can be predicted.\footnote{This is
evident from the formulae (\ref{fourone}) and (\ref{fourthree}) and has
been widely discussed in quantum mechanics in various contexts.  See
\cite{Unr86,AV91,GH90a} for recent examples.}  Additional information
about the existence of future non-chronal regions summarized by $\rho_f$ in
(\ref{fourtwo}) is required. There is thus no notion of the state of the
system on a spacelike surface in this generalized quantum mechanics.

The existence of non-unitary evolution in the future not only acausally
affects the probabilities of present alternatives, it also
affects their decoherence.  Consider, for example, a set of histories
defined by alternatives $\{P_\alpha(\sigma)\}$ a single spacelike
surface that is before any non-chronal region.  The decoherence functional
according to (\ref{threethirteen}) is
\begin{equation}
D(\alpha^\prime,\alpha) = N\ Tr \bigl[XP_{\alpha^\prime} (\sigma) \rho
P_\alpha (\sigma) X^\dagger\bigr]\ .
\label{fourseven}
\end{equation}
Were
$X$  unitary, {\it any}~set of alternatives {\it automatically}
decoheres because of the cyclic property of the trace.  If $X$ is
non-unitary  then only certain sets of $P_\alpha$ will decohere.
Decoherence is therefore acausally affected by the
spacetime geometry of the future.  However, typical mechanisms of
decoherence that involve the rapid dispersal of phase information among
ignored variables that interact with those of interest operate
essentially locally in time.\footnote{See, for example, the discussion
in \cite{GH90a} and the references therein.}
  Such mechanisms may be essentially
unaffected by non-unitary evolution in the future.  We may, for example,
continue to expect the decoherence of alternatives that define the
present quasiclassical domain of familiar experience even in the
presence of a modest number of future non-chronal regions.

As we have seen,
generalized quantum mechanics with non-unitary evolution
violates causality because information about the
future is required to calculate the probabilities in the present.
However, it is important to stress that it is not information about the
specific alternatives that occur in the future that is required.  Rather
it is information about the future geometry of spacetime that enters as
a fixed input in this quantum field theory in curved spacetime.
Independence of present probabilities on
specific alternatives that occur in the future is guaranteed by the
probability sum rules that follow from decoherence.  If $\{\alpha\}$
denotes a set of alternatives accessible to us and $\{\beta\}$ another
set in the future such that $\{\alpha, \beta\}$ decoheres, then the
probability of alternative $\alpha$ can be calculated two
ways.  First it can be calculated directly from (\ref{fourone}). Second, it
can be calculated by saying that one of the alternatives
$\{\beta\}$ occurs in the future and summing $p(\alpha, \beta)$ over the
unknown values of $\{\beta\}$.  These two calculations agree because the
alternatives
$\{\alpha, \beta\}$ decohere, [\cf (\ref{threefour})]
\begin{equation}
p(\alpha) \approx \sum_\beta p (\alpha, \beta)\ . \label{foureight}
\end{equation}
For example\footnote{Suggested to the author by J.~Friedman},
 suppose a non-chronal region of spacetime  exists in the
future but is contained inside an impenetrable box with a door.
Observers in the future have the alternatives of opening the
door to let fields propagate in this region or leaving it closed and
preventing fields from interacting with it. Present probabilities  are
affected by the existence of such a region, whether the door is opened
or not, but unaffected by the specific decision the future observers
take.

In a quantum mechanics based on the decoherence of coarse-grained sets
of alternative histories, probability sum rules like (\ref{foureight}) hold in
much wider circumstances than those described above.  For example, the
probabilities of alternatives $\{\alpha\}$ in the present are
independent of specific unknown alternatives $\{\beta\}$ whether these
are in the future, past or spacelike separated from $\{\alpha\}$
provided the joint set of alternatives decoheres.  Even in a quantum
theory of gravity where spacetime geometry is a dynamical variable, and
we could envision observers in the future deciding whether to create
non-chronal regions or not, similar results would be expected to hold.
Thus we could not
determine by present observations whether observers in the future will
decide to build Morris, Thorne, Yurtsever-type time machines.

\section{Testing Non-Unitary Evolution and Causality Violation}
\label{sec:five}

How might the non-unitary evolution and causality violation of the
present generalized quantum mechanics of non-chronal spacetimes be
tested in the laboratory?  This section offers a preliminary discussion.

We begin with a simple model of scattering through a non-chronal region
of spacetime.  More specifically, we imagine that a small non-chronal
region of spacetime has been located and that we direct particle beams
so as to interact in that region, measuring their asymptotic
states by apparatus that does not itself interact with the non-chronal
region to a good approximation.  The incoming particles are prepared in
pure initial states and final pure states are detected by the apparatus.
This is certainly not the most general measurement situation that can be
envisioned but gives a simple illustration of the effects of
non-unitarity.\footnote{A more general measurement theory may be
exhibited along the lines described in Ref.~\cite{Har91a}, Section
II.10.  It is subject to limitations of the character described in
Ref.~\cite{Har91b} regarding the influence of the measuring apparatus on
the probabilities of the measured alternatives.}

We suppose the Hilbert space of the closed system factors into a
tensor product, ${\cal H}_s\otimes$ ${\cal H}_r$ of a Hilbert space ${\cal
H}_s$ describing the scattering particles and a Hilbert space ${\cal H}_r$
describing the rest, including the apparatus for preparation and
detection.  We consider an initial state $\rho$ at time $t_1$
corresponding to the preparation of the experiment as described above.  We
consider the alternatives for the scattering particles in which they are
prepared and detected  in members
of  complete sets of states $\{|\alpha, t\rangle\}$ in ${\cal H}_s$ at
various times (\eg wave packets with approximately definite momentum).
These alternatives are represented as a set of projection
operators
\begin{equation}
S_\alpha(t) = |\alpha, t\rangle\langle\alpha, t|\otimes I_r
\label{fiveone}
\end{equation}
in the Heisenberg-like picture represented in Section III.

The histories describing the scattering process are represented
by
the chains $S_\beta(t_2) S_\alpha (t_1)$ where $t_1$ and $t_2$ are the
initial and final times of the scattering process.  The decoherence
functional for these histories is then [\cf (\ref{threenine})]
\begin{equation}
D\left(\beta^\prime, \alpha^\prime ; \beta\alpha\right)
= N\, Tr\left[S_{\beta^\prime} (t_2) X S_{\alpha^\prime} (t_1) \rho
S_\alpha (t_1) X^\dagger S_\beta (t_2)\right]
\label{fivetwo}
\end{equation}
where $N$ is given by (\ref{threesix}).

The defining assumption of the model is that, with an initial
$\rho$ appropriate to the experimental set-up described above, the
non-unitary evolution operator $X$ effectively acts only on ${\cal
H}_s$. Noting that the decoherence of the alternatives $\beta$ at time
$t_2$ is automatic because of the cyclic property of the trace, we may
then write
\begin{equation}
D\left(\beta^\prime, \alpha^\prime ; \beta, \alpha\right) =
\delta_{\beta^\prime\beta} N\, \left\langle\beta, t_2 | X |
\alpha^\prime, t_1\right\rangle \left\langle\alpha^\prime, t_1|Sp (\rho)
 | \alpha, t_1
\right\rangle\left\langle\alpha, t_1|X^\dagger | \beta, t_2\right\rangle
\label{fivethree}
\end{equation}
where $Sp(\rho)$ is the operator on ${\cal H}_s$ that is the trace of
$\rho$ over ${\cal H}_r$ and $X$ is the restriction of the
non-unitary evolution to ${\cal H}_s$.

The decoherence of the measured alternatives $\alpha$ at the initial
time $t_1$ is not automatic.  However,  in this model of a measurement
we {\it assume} that the interaction of the particles
with the apparatus effects the
decoherence of the alternative initial states of the particles, say, by
correlation with an independent, persistent record of the their
 initial state
(as in Ref.~\cite{Har91a}, Section II.10).  Effectively we assume
\begin{equation}
\left\langle\alpha^\prime, t_1 | Sp(\rho) | \alpha, t_1 \right\rangle
\propto \delta_{\alpha^\prime\alpha}\ .
\label{fivefour}
\end{equation}
The joint probabilities for this now decoherent set of histories are
\begin{equation}
p(\beta, \alpha) = N| \left\langle \beta, t_2 | X | \alpha, t_1 \right\rangle
 |^2
\left\langle\alpha, t_1 | Sp (\rho) | \alpha, t_1\right\rangle\ .
\label{fivefive}
\end{equation}

In scattering experiments it is not so much the joint probability
$p(\beta, \alpha)$ of both initial and final states that is of interest, but
rather the conditional probability $p(\beta|\alpha)$ of a final state, $\beta$,
given an initial one, $\alpha$.  This is constructed in the standard
way [\cf (\ref{fourfour})]
with the result
\begin{equation}
p(\beta|\alpha) = \frac{|\left\langle\beta, t_2 | X | \alpha, t_1
\right\rangle |^2}{\Sigma_\beta | \left\langle\beta, t_2 | X |\alpha,
t_1\right\rangle|^2}\ .
\label{fivesix}
\end{equation}
All reference to the external apparatus has canceled from this effective
expression for conditional probabilities.  Except for the normalizing
denominator, (\ref{fivesix}) is the usual expression the probability of a
scattering process. Indeed, were $X$ unitary, the denominator would be
unity.  The net effect of the non-unitarity of $X$ has simply been to
normalize the usual quantity $|\langle\beta t_2| X | \alpha
t_1\rangle|^2$ so that the probability sum rule
$\Sigma_\beta p (\beta|\alpha) = 1$ is satisfied.

Were small size
 non-chronal regions widespread in spacetime,
the difference between the probabilities predicted by (\ref{fivesix})
and a standard formula for the same scattering in flat spacetime would
be both a means of detecting such non-chronal regions and verifying the
non-unitarity of evolution through them.  In the absence of
estimates of the sizes and density of non-chronal regions and of the
$X$'s which describe the evolution through them, we cannot provide
estimates of the effect of non-unitary evolution here.  It is
through (\ref{fivesix}), however, that such effects would be calculated
\cite{Rosup}.

In a similar way, we could estimate the acausal effects of future
non-chronal regions on present experiments.  We would compare standard
flat space formulae for probabilities with those computed from formulae
like (\ref{threethirteen}) with a non-unitary $X$ describing the effect
of non-chronal regions in the future on present measurement situations.
We would then be led to formulae like
(\ref{fivethree}) or (\ref{fivesix}) for the probabilities of the
measured subsystem but with the non-unitary $X$ to the {\it future} of
the projections describing the measurement.  The measurement situation
would have a chance of detecting a departure from strict causality only
if there were a significant probability that the subsystem under study
interacted with a non-chronal region subsequent to the
time of the experiment.  If non-chronal regions are sparse in the
future history of spacetime, then we might expect this probability
to be very,
very small and the resulting violation of causality negligible.  If,
however, there were a roiling sea of non-chronal regions near a generic
final singularity, then the probabilities for causality violation might
be more interesting.  The present generalized quantum mechanics provides
a way of estimating them.

\section{Information}
\label{sec:info}

In usual quantum mechanics, the state of a system on a spacelike surface
$\sigma$ is as complete a description of that system as it is possible
to give.  When the state is given, the missing information about the
system is zero.  When it is only possible to give
probabilities for an ensemble
of possible states, the system is described by a (Schr\"odinger-picture)
density matrix $\rho_\eff(\sigma)$.  The missing information is then given
more generally by
\begin{equation}
S(\sigma) = -Tr \bigl[\rho_\eff(\sigma) \log \rho_\eff (\sigma)\bigr]
\label{sixone}
\end{equation}
which vanishes when the probability is unity for a single state and zero
for all orthogonal ones, \ie when $\rho$ is pure.  This missing
information is
conserved under unitary evolution.

The generalized quantum mechanics of matter fields in non-chronal
spacetimes that is
under discussion in this paper does not generally permit a
notion of ``state on a spacelike surface'' as discussed in Section IV.
How then do we define the
information available about these fields on a spacelike surface?  How
does this information evolve? This Section is devoted to some answers to
these questions.

As in the discussion of Sections III and IV, we begin by considering
a spacetime with
a single non-chronal region.  We will consider two spacelike surfaces
$\sigma^\prime$ and $\sigma^{''}$ the first of which is {\it before} the
non-chronal region and the second {\it after} (Figure 1).
On these surfaces we consider
{\it decoherent} sets of alternatives represented by sets of orthogonal,
commuting projection operators $\{P^\prime_\beta\}$ and $\{P^{''}_\beta\}$
respectively.  The probabilities of these alternatives are given by the
diagonal elements of the decoherence functional, specifically
[\cf (\ref{threethirteen}) and (\ref{threefourteen})]
\begin{mathletters}
\label{sixtwo}
\begin{eqnarray}
p\left(\alpha; \sigma^\prime, \{P^\prime_\beta\}\right)&
 = N\, Tr\left(XP^\prime_\alpha \rho\, P^\prime_\alpha X^\dagger\right)\ ,
\label{mlett:sixtwo a}\\
p\left(\alpha; \sigma^{\prime\prime}, \left\{P^{\prime\prime}_\beta\right\}
\right)
& = N\, Tr\left(P^{\prime\prime}_\alpha X \rho X^\dagger
P^{\prime\prime}_\alpha\right)\ . \label{mlett:sixtwo b}
\end{eqnarray}
\end{mathletters}

The assumed decoherence of the alternatives allows the probabilities
(\ref{sixtwo})
to be written in other useful ways.  When (\ref{threeone}) is satisfied
\begin{equation}
p(\alpha) \equiv D(\alpha, \alpha) \approx \sum_{\alpha^\prime} D
(\alpha^\prime, \alpha) \approx \sum_{\alpha^\prime} D (\alpha,
\alpha^\prime) \ . \label{sixthree}
\end{equation}
yielding the expressions
\begin{mathletters}
\begin{eqnarray}
p\left(\alpha; \sigma^\prime, \left\{P^\prime_\beta\right\}\right)
& \approx N\, Tr\left(P^\prime_\alpha \rho X^\dagger X\right)
&\approx N\, Tr \left(P^\prime_\alpha X^\dagger X\rho\right)
\label{mlett:sixfour a}\\
p\left(\alpha; \sigma^{\prime\prime},
\left\{P^{\prime\prime}_\beta\right\}\right)
&\approx N\, Tr \left(P^{\prime\prime}_\alpha X\rho X^\dagger\right)\ .
\label{mlett:sixfour b}
\end{eqnarray}
\end{mathletters}
These relations may be written more symmetrically as
\begin{mathletters}
\begin{eqnarray}
p\left(\alpha; \sigma^\prime, \left\{P^\prime_\beta\right\}\right)
& \approx Tr\left(P^\prime_\alpha\hat\rho\right)\ ,
\label{mlett:sixfive a}\\
p\left(\alpha; \sigma^{\prime\prime},
\left\{P^{\prime\prime}_\beta\right\}\right)
&\approx Tr\left(P^{\prime\prime}_\alpha\bar\rho\right)\ ,
\label{mlett:sixfive b}\\
Tr \left(\rho\bigl[X^\dagger X, \rho\bigr]\right)
&\approx 0\ , ~~~~~~~~~~
\label{mlett:sixfive c}
\end{eqnarray}
\end{mathletters}
where $\hat\rho$ and $\bar\rho$ are the density matrices
\begin{eqnarray}
\hat\rho & =& \bigl\{\rho, X^\dagger
X\bigr\}/Tr\bigl(X\rho X^\dagger\bigr)\ ,
 \label{sixsix}\\
\bar\rho & =& X\rho X^\dagger/Tr\bigl(X\rho X^\dagger\bigr)\ ,
\label{sixseven}
\end{eqnarray}
$\{\cdot , \cdot \}$ denoting an anticommutator.
 These relations will be helpful in what follows.

With these preliminaries we turn to a definition of the missing
information on a spacelike surface $\sigma$. First, generalizing a
construction of Jaynes \cite{Ros83,GH90a},
we define, $S(\sigma; \{P_\beta\})$, the
missing information on $\sigma$ relative to a set of alternatives
$\{P_\beta\}$ whose probabilities are $\{p_\beta\}$.  This is the
maximum of the entropy functional
\begin{equation}
{\cal S} (\tilde\rho) = -Tr\left(\tilde\rho \log \tilde\rho\right)
\label{sixeight}
\end{equation}
over all density matrices $\tilde\rho$ that reproduce the probabilities
$p_\alpha$ through $Tr(P_\alpha\tilde\rho)=p_\alpha$.  Of course, it is
possible to lose information about a system by asking stupid questions.
To obtain a good measure of the missing information on a spacelike surface,
$S(\sigma)$, we, therefore, minimize $S(\sigma; \{P_\beta\})$ over all
decohering sets of alternatives $\{P_\beta\}$.  Explicitly then,
\begin{equation}
S(\sigma) = \mathop{\min}_{\buildrel {\rm decoherent}
\over {\{P_\alpha\}}}
\left\{\matrix{\mathop{\max}\limits_{\buildrel \tilde \rho
\ {\rm with} \over {Tr\left(P_\alpha \tilde\rho\right)=p_\alpha}}&
{\cal S}(\tilde\rho)\cr}\right\}\ .
\label{sixnine}
\end{equation}

If we assume for a moment that the projections $\{P_\alpha\}$ are all
onto finite dimensional subspaces of the Hilbert space ${\cal H}$,
then the
density matrix that maximizes ${\cal S}(\tilde\rho)$ subject to the
probability constraint is a standard construction
\begin{equation}
\tilde\rho_{\rm max}  = \sum_\alpha\left[p(\alpha; \sigma, \{P_\beta\})
\, P_\alpha/Tr(P_\alpha)\right]\ ,
\label{sixten}
\end{equation}
and
\begin{equation}
S(\sigma; \{P_\beta\}) = {\cal S}(\tilde\rho_{\max})
 = \sum_\alpha p(\alpha; \sigma, \{P_\beta\}) \log
\left[p(\alpha; \sigma, \{P_\beta\})/Tr (P_\alpha)\right]
\ .
\label{sixeleven}
\end{equation}
The limit of $\{P_\alpha\}$ with infinite trace can then be considered.

Eq.~(\ref{sixnine}) defines the missing information a spacelike surface
$\sigma$  and holds whether the probabilities $\{p_\alpha\}$ for the
alternatives $\{P_\alpha\}$ follow from a
notion of state on the spacelike surface, as in usual quantum mechanics,
or from a generalized quantum mechanics decoherence functional as in
(\ref{threethirteen}). It is instructive to see how the familiar result
(\ref{sixone}), is a
consequence of this definition when the probabilities $p_\alpha$ {\it are}
given by a density matrix $\rho_\eff(\sigma)$ through
\begin{equation}
p_\alpha = Tr\left[P_\alpha \rho_\eff (\sigma)\right]\ .
\label{sixtwelve}
\end{equation}
To see this note that from (\ref{sixten}), that when (\ref{sixtwelve}) holds
\begin{equation}
-Tr\left[\tilde\rho_{\rm max} \log \tilde\rho_{\rm max}\right] =
-Tr\left[\rho_\eff(\sigma) \log \tilde\rho_{\rm max}\right]\ .
\label{sixthirteen}
\end{equation}
The inequality\footnote{For a convenient proof, see \cite{Rue69} or the
discussion in the Appendix.}
\begin{equation}
-Tr\left(\rho_1 \log \rho_2\right) \geq - Tr \left(\rho_1 \log
\rho_1\right)\ ,
\label{sixfourteen}
\end{equation}
which holds for any two density matrices $\rho_1$ and $\rho_2$, then
implies
\begin{equation}
S\left(\alpha; \left\{P_\alpha\right\}\right) \equiv {\cal S}
\left(\tilde\rho_{\rm max}\right) \geq {\cal S}\, [\rho_\eff(\sigma)]\ .
\label{sixfifteen}
\end{equation}
The lower bound is reached by choosing the projections $\{P_\alpha\}$ to
be onto a basis in which $\rho_\eff(\sigma)$ is diagonal.
In this connection
the $Tr(P_\alpha)$ term in (\ref{sixeleven}) which contributes
$+\Sigma_\alpha p(\alpha) \log\, Tr(P_\alpha)$ to the missing
information is important.  It is that term which favors choosing
$P_\alpha$ which are as refined (low dimensional) as possible.
 In usual quantum
mechanics, the decoherence functional for alternatives defined on a single
spacelike surface is $D(\alpha^\prime, \alpha) =
Tr[P_{\alpha^\prime}\rho_\eff(\sigma)P_\alpha]$, so that {\it any} set of
alternatives $\{P_\alpha\}$ decoheres because of the cyclic property of
the trace.  When the minimization of $S(\sigma; \{P_\alpha\})$ over
decoherent alternatives is carried out in the definition (\ref{sixnine}),
we therefore find
\begin{equation}
S(\sigma) = \mathop{\min}\limits_{\{P_\beta\}} S\left(\sigma;
\left\{P_\beta\right\}\right) = -Tr\left[\rho_\eff(\sigma) \log \rho_\eff
(\sigma)
\right]
\label{sixsixteen}
\end{equation}
--- the familiar and anticipated result.

We now return to the question of the relation between the missing
information $S(\sigma^\prime)$ and
$S\left(\sigma^{\prime\prime}\right)$ on two spacelike surfaces before
and after a non-chronal region in the generalized quantum theories of
Section III.  Eq.~(\ref{threenine}) shows that the decoherence functional
for alternatives {\it after} the last non-chronal region is the same as
that of usual quantum mechanics with the effective density matrix
$\bar\rho$ defined in (\ref{sixseven}).  In particular, the cyclic
property of the trace shows that alternatives confined to
a single surface after
all non-chronal regions always decohere.  We therefore have from the
argument in the preceding paragraph
\begin{equation}
S\left(\sigma^{\prime\prime}\right) = {\cal S}\left(\bar\rho\right)
= {\cal S} \left(\frac{X\rho X^\dagger}{Tr\left(X\rho X^\dagger\right)}\right)
\ .\label{sixseventeen}
\end{equation}
The missing information (\ref{sixseventeen})
 bears no special relation to the missing information in $\rho$ that
would be calculated in usual quantum mechanics according to
(\ref{sixone}).  Indeed, one can show that, for any $\rho$, there is
always an operator $X$ that will give ${\cal S}(\bar\rho)$ any value
from zero up to its maximum.  That fact, however, is of no special
interest in generalized quantum mechanics.  Both $X$ and $\rho$ are
needed in the computation of all probabilities and there are thus no
alternatives for which $S(\sigma)$ is given by (\ref{threeone}).

The missing information, $S(\sigma^\prime)$ on a surface $\sigma^\prime$
{\it before} a non-chronal region is not as easily calculable as
$S(\sigma^{\prime\prime})$ because the strictures of decoherence are
non-trivial.  However, it is possible to show that $S(\sigma^\prime)$
cannot be less than $S(\sigma^{\prime\prime})$.  The demonstration
proceeds along the lines of the evaluation of $S(\sigma)$ for usual
quantum mechanics as derived above, but requires a more general inequality
than (\ref{sixfourteen}). Specifically, it requires the inequality
\begin{equation}
-Tr (A \log \rho_2) \geq - Tr (A \log A) \label{sixeighteen}
\end{equation}
valid for any density matrix $\rho_2$ and  linear operator $A$ of the form
\begin{equation}
A=B\rho_1 B^{-1}\ ,
\label{sixnineteen}
\end{equation}
for some $B$ and density matrix $\rho_1$,
such that the left-hand-side of (\ref{sixeighteen}) is real and
positive. The condition
(\ref{sixnineteen}) is enough to allow the definition of $\log A$.  This and
the derivation of (\ref{sixeighteen}) are discussed in the Appendix.

We can apply this inequality to obtain a lower bound on
$S(\sigma^\prime; \{P_\beta\})$ as given by (\ref{sixeleven}) by noting
that, according to (\ref{mlett:sixfour a}), the probabilities for
decohering
alternatives $\{P_\alpha\}$ on $\sigma^\prime$ may be written
\begin{equation}
p\left(\alpha; \sigma^\prime, \left\{P_\beta\right\}\right) =
Tr\left(P_\alpha A\right)
\label{sixtwenty}
\end{equation}
where $A$ is the operator
\begin{equation}
A=\frac{X^\dagger X \rho}{Tr\left(X^\dagger X\rho\right)}\ .
\label{sixtwentyone}
\end{equation}
The operator $A$ is not a density matrix. (It is not generally even
Hermitian.) However, it is of the form (\ref{sixnineteen}) with
$B=\left(X^\dagger\right)^{-1}$ and $\rho_1 = \bar\rho = X\rho X^\dagger
/Tr(X\rho
X^\dagger)$.  It is an elementary calculation in a basis in which
$\tilde\rho_{\rm max}$ of (\ref{sixten}) is diagonal to verify
\begin{equation}
-Tr\left(A \log \tilde\rho_{\rm max}\right) = -Tr \left(\tilde\rho_{\rm
max} \log \tilde\rho_{\rm max}\right)\label{sixtwentytwo}
\end{equation}
We may use this and (\ref{sixeighteen}) with $\rho_2 = \tilde\rho_{max}$
to derive
\begin{equation}
S\left(\sigma; \left\{P_\beta\right\}\right) \equiv {\cal S}\left(\tilde
\rho_{\rm max}\right) \geq - Tr\bigl(A \log A\bigr)\ .
\label{sixtwentythree}
\end{equation}
However, the cyclic property of the trace together with the definition
of $\log A$ (see Appendix) imply that $Tr(A \log A) = Tr (\bar\rho
\log \bar \rho)$ which is $S(\sigma^{\prime\prime})$ [\cf
(\ref{sixseventeen})]. When the minimum of the left-hand side of
(\ref{sixtwentythree}) is taken over all decoherent sets $\{P_\beta\}$
on $\sigma^\prime$, we therefore have
\begin{equation}
S(\sigma^\prime) \geq S\left(\sigma^{\prime\prime}\right)
= {\cal S}(\bar\rho) =
{\cal S} \left(\frac{X\rho X^\dagger}{Tr\left(X\rho X^\dagger
\right)}\right)\ .
\label{sixtwentyfour}
\end{equation}
Thus, information can be gained but not lost in evolving from a
spacelike surface before a non-chronal region to a spacelike surface
after all such regions.

The inequality (\ref{sixtwentyfour}) becomes an equality in the special
case that $X^\dagger X$ commutes with $\rho$.  Then the projections onto a
basis in which $\rho$ and $X^\dagger X$ are simultaneously diagonal are
decoherent and give $\tilde\rho_{\rm max} = \bar\rho$.  If $X^\dagger X$
and $\rho$ do not commute, then there certainly is no set of
one-dimensional projections that are decoherent, so one would conjecture
that the bound provided by ${\cal S}(\bar\rho)$ is not realized.  The
determination of an optimum lower bound on $S(\sigma^\prime)$ then
becomes an interesting question.

The possibility of information gain in moving from one spacelike surface
is not surprising from the point of view of the requirements of
decoherence.  On a spacelike surface $\sigma^{\prime\prime}$ to the
future of {\it all} non-chronal regions {\it any} set of alternatives
$\{P_\beta\}$ is decoherent.  On a surface $\sigma^\prime$ to the past
of {\it some} non-chronal regions only certain sets of alternatives will
decohere.  There are thus more questions with which to extract
information about the quantum system on $\sigma^{\prime\prime}$ than on
$\sigma^\prime$ and a corresponding decrease in missing information is
to be expected.

We have so far considered the case of a spacetime with a single
non-chronal region, but a general spacetime may have many.  When
spacelike surfaces may be passed between such regions, we may consider
the missing information on each and the evolution of the missing
information from surface to surface.  In general, we do not expect any
particular relation between the values of $S(\sigma)$ as we pass from
one surface to a later one.  It may increase or it may decrease
depending on the requirements of decoherence.  However, on each surface
the
missing information must satisfy the inequality (\ref{sixtwentyfour})
where $\sigma^{\prime\prime}$ is to the future of all non-chronal
regions.  On such a surface the strictures of decoherence are least and
the information available about the system the most.

As we discussed in the Introduction, generalized quantum mechanics
permits quantum theory to be formulated in fully four-dimensional form
that does not rely on a notion of state on a spacelike surface.  In such
a theory complete information about a system is not necessarily
available on every spacelike surface.  Rather the appropriate notion of
information is itself four-dimensional as described in this Section.  It
may be necessary to search in many regions of spacetime to recover
complete information about a system \cite{Harup}.

\section{No Non-Conservation of Energy, Signaling Faster than Light, or
Everett Phones}
\label{sec:seven}

Generalized quantum mechanics is a modest generalization of familiar
quantum theory that retains the principle of the linear superposition of
amplitudes in the form (\ref{threethree}).  Generalized quantum
theories, such as the one under discussion, may have a notion of a
Heisenberg-like state that specifies the initial and final conditions,
but will not always permit a notion of an evolving state on a spacelike
surface.  Various other generalizations of quantum mechanics have been
proposed that retain the notion of a state on a spacelike surface but
abandon or modify the principle of superposition in some way.  Recent
examples, are the work of Banks, Peskin, and Susskind \cite{BPS84} and
Srednicki \cite{SreXX} in which pure density matrices evolve into mixed
ones, and Weinberg's non-linear quantum mechanics \cite{Wei89}.  The
generalization of Banks, Peskin, and Susskind suffers from energy
non-conservation while that of Weinberg can permit communication faster
than light and communication with
alternative branches of the universe in situations that have been called
the ``Everett phone'' by  Polchinski \cite{Pol91}.

The above generalizations of quantum mechanics are non-linear because
they incorporate a non-linear law of evolution for states on a spacelike
surface.  The generalized quantum mechanics for non-chronal spacetimes
under discussion in this paper cannot be characterized as linear or
non-linear in this way because it does not generally permit a notion of
state on a spacelike surface much less a discussion of the law for its
evolution. This generalization does respect the linear principle of
superposition in the sense of (\ref{threethree}).  However, because of
the normalization factor (\ref{threesix}) probabilities are not
quadratically related to a pure state vector describing the initial
condition as they would be in usual quantum mechanics.  For this reason
it is prudent to examine the present generalized quantum mechanics
for energy non-conservation, signaling faster than light, and Everett
phones.  In this Section we shall show that decoherence prohibits all of
these anomalies. Our arguments apply to all generalized quantum
theories although we shall describe them here for the particular case
of the generalized quantum mechanics of fields in non-chronal
spacetimes.

\noindent {\bf Energy Conservation}

When spacetime geometry is
 time-dependent we do not expect conservation of the total energy of
matter fields moving in it, even classically.  Where
there are compact non-chronal regions of spacetime, the geometry will
certainly be time dependent.  However, we can still analyse the question of
energy conservation in those regions where spacetime {\it is} locally
time independent.
Specifically, consider a region of spacetime that is foliable by
spacelike surfaces labeled by a co\"ordinate $t$ such that
$\partial/\partial t$ is a Killing vector
which asymptotically corresponds to a time-translation in some
Lorentz frame.  We can then define the energy-momentum four-vector of
the matter fields on a spacelike surface of constant $t$ as
\begin{equation}
P^\alpha (t)
= \int_t d\Sigma^\beta T^\alpha_\beta
\label{sevenone}
\end{equation}
where $T^\alpha_\beta$ is the stress-energy of the matter fields and
$d\Sigma^\beta$ is an element of the surface of constant $t$. In particular
$P^t\equiv H$ is the total energy of the matter fields and the corresponding
quantum mechanical operator is the generator of translations in $t$. The
total energy is conserved between surfaces of constant $t$ because
$\partial/\partial t$ is a Killing vector, which means that the
operator $H$ is independent of time.

Whether energy is conserved quantum mechanically is a question of the
probabilities for the correlation of the values of $H$ on two different
surfaces of constant $t$.  A specific example of the calculation of such
probabilities will illustrate all the features of the general case.
Consider a single non-chronal region as discussed in Section III, and
suppose that before the non-chronal region there is a region of spacetime
with a time-translation symmetry in the sense discussed above.
Let $\{P^H_\alpha (t)\}$ denote a set of projections onto an exhaustive
set of ranges $\{\Delta_\alpha\}$ of the total energy in matter fields,
$H$, in the Heisenberg-like picture specified by (\ref{threesevena}).
Since $H$ is independent of $t$ before the non-chronal region, the
projections $\{P^H_\alpha (t)\}$ are also.

Now consider a set of histories which contain the projections
$\{P^H_\alpha (t)\}$ at two different times $t_1$ and $t_2$ in the
region of time-translation
 symmetry.  The chain of projections before $\sigma_-$ [\cf
(\ref{threeten})] would have the form
\begin{equation}
C_\alpha = C^c_{\alpha_c} P^H_{\alpha_2} (t_2) C^b_{\alpha_b}
P^H_{\alpha_1} (t_1) C^a_{\alpha_a}
\label{seventwo}
\end{equation}
where $C^a_{\alpha_a}, C^b_{\alpha_b}$, and $C^c_{\alpha_c}$ are
themselves chains of projections.  Suppose this set of alternative
histories decoheres.  The joint probabilities for the individual
histories may be calculated from (\ref{threeone}) and
(\ref{threetwelve}).  Conservation of energy would mean
\begin{equation}
p\left(\beta, \alpha_c, \alpha_2, \alpha_b, \alpha_1, \alpha_a\right)
\propto \delta_{\alpha_1\alpha_2}
\label{seventhree}
\end{equation}
for any choice of the other alternatives $\alpha_a, \alpha_b, \alpha_c$
and $\beta$.  Eq.~(\ref{seventhree}) is not
a consequence of any operator identity since $C^b_{\alpha_b}$ in
(\ref{seventwo}) need not commute with $H$.  However, it is a
consequence of decoherence.\footnote{The argument appears to be part of
the lore of consistent histories.  The author learned it from
R.W.~Griffiths.  For a more detailed discussion including a
consideration of approximate decoherence see \cite{LHXX}.} Decoherence
guarantees the consistency of probability sum rules.  Thus, in
particular
\begin{equation}
p\left(\alpha_2, \alpha_1\right) = \sum_{\beta, \alpha_c, \alpha_b,
\alpha_a} p\left(\beta, \alpha_c, \alpha_2, \alpha_b, \alpha_1, \alpha_a
\right)\ . \label{sevenfour}
\end{equation}
However, the history which consists just of alternative values of the
energy at time $t_1$ and $t_2$ is represented by the chain
\begin{equation}
P^H_{\alpha_2} (t_2) P^H_{\alpha_1} (t_1) \propto
\delta_{\alpha_2\alpha_1}\ .
\label{sevenfive}
\end{equation}
This chain vanishes unless $\alpha_1 = \alpha_2$ because the projections
onto the values of a conserved quantity
are independent of time and projections for
different alternatives are orthogonal.  Thus, $p(\alpha_2,
\alpha_1) \propto \delta_{\alpha_1\alpha_2}$. Since the right-hand side
of (\ref{sevenfour}) is a sum of positive numbers, (\ref{seventhree})
follows also.

Obvious extensions of this argument show that, in general, decoherence
guarantees the conservation of energy in regimes of spacetime that
possess a time-translation symmetry in the sense described above.

\noindent {\bf No Signaling Faster than Light}

The meaning of ``signaling faster than light'' requires careful
definition.  It is not simply a matter of being able to infer
the probabilities of
alternatives in one spacetime region from alternatives
in a spacelike separated
 region. The EPRB situation in
which two spins prepared in a singlet state move into separate spacelike
regions is an example.
 From a determination of the spin direction of one it is
possible to infer the spin of the other.  That, however, is not a signal
sent faster than light.  The determination only exploits
 a correlation present in the initial state.  To investigate whether it is
possible to signal faster than light we should investigate whether from
alternatives in one spacetime region one can infer the probabilities of
another system in another spacelike separated region when the two
systems
were {\it initially uncorrelated} and remained spacelike
separated at all subsequent times.

A classic example in the EPRB situation is the question of whether
carrying out a measurement on one spin can influence the probabilities
of the second.  From a closed system point of view there is a third
system --- the apparatus --- which is initially uncorrelated with either
spin and which
 subsequently becomes correlated with one of them with a certain
probability while the remaining spin remains always
spacelike separated. As we shall
show below it is
straightforward to show from the causality of field theory that
spacelike separated
alternatives remain uncorrelated if they were initially uncorrelated and
there is no signaling faster than light.

When spacetime possesses non-chronal regions, correlations arise not only
from an initial condition but also
from future non-chronal regions.  We remarked in Section
IV that the formula (\ref{fourthree}) for the probabilities of
alternatives to the past of all non-chronal regions was like that of
quantum mechanics with a final as well as an initial condition.  Systems
that are now
spacelike separated may be in causal contact both in the past and
future and correlations arising from both initial and final conditions
could exist in the present between them.  We
should count neither as signaling faster than light.  To meaningfully
consider the possibility of signaling faster than light, we should
consider two systems that are uncorrelated both  with respect to the
initial condition of the system and also any conditions that may exist in
the future.

The situation may be illustrated with a simple example.  Consider two
spacelike separated regions $R_1$ and $R_2$ whose pasts
intersect the initial
surface in disjoint regions $D_1$ and $D_2$. Suppose that the fields
in $D_1$ and $D_2$ are initially uncorrelated.  This means that the
initial state can be written as $\rho_1 \otimes \rho_2$ where the fields
in $D_1$ act only on $\rho_1$ and the fields in $D_2$ act only on
$\rho_2$.

Similarly, suppose that there are no non-chronal regions to the past of
either $R_1$ or $R_2$ and no non-chronal regions in the common future of
both $D_1$ and $D_2$.  In particular, this means that there are no
common
non-chronal regions in the future of both $R_1$ and $R_2$ and ensures
that there are no correlations between fields in these regions by virtue of
future non-chronal regions.

In a causal field theory the field operators in a spacetime region $R$
are related by the Heisenberg equations of motion only to operators in
the future and past of that region.  The non-unitary evolution $X$ thus
factors into a part $X_1$ acting only on $\rho_1$, referring to the
non-chronal regions in the future of $D_1$, and a similar $X_2$ acting on
$\rho_2$.  The joint probability of a decoherent set of alternatives
$P_{\alpha_1} (R_1)$ referring to region $R_1$ and another
$P_{\alpha_2}(R_2)$ referring to $R_2$ then also factors
\begin{equation}
p\left(\alpha_1, \alpha_2\right) = N_1 Tr_1\left[X^\dagger_1 X_1
P_{\alpha_1} (R_1) \rho_1\right]\, N_2 Tr_2 \left[X^\dagger_2 X_2
P_{\alpha_2} (R_2) \rho_2 \right]\ .
\label{sevensix}
\end{equation}
where $N_1 = Tr_1(X_1\rho_1X^\dagger_1)$ and
$N_2=Tr_2(X_2\rho_2X^\dagger_2)$.  The causality of field theory
together with the absence of initial and final correlations thus implies
the factorization of joint probabilities of alternatives in spacelike
separated regions
\begin{equation}
p\left(\alpha_1, \alpha_2\right) = p\left(\alpha_1\right) p
\left(\alpha_2\right)\ .
\label{sevenseven}
\end{equation}
The probabilities of alternatives in spacelike separated regions are
thus independent, and signaling faster than light is not possible.

The fact that it is not possible to signal faster than light in the
absence of built in correlations does not mean that it would not be
interesting to investigate exploiting
 the correlations between spacelike separated
regions provided by non-chronal regions in their common future.

\noindent{\bf Everett Phones}

In his analysis of Weinberg's non-linear quantum mechanics, Polchinski
\cite{Pol91}
has given an example of a kind of ``communication'' between
different branches of the wave function that he dubbed the ``Everett
phone''. Specifically, if briefly, he considers sets of histories of a
spin--$\half$ ion in a Stern-Gerlach apparatus and a ``macroscopic
observer''. At time $t_1$, the $z$-component of the spin is determined by
the splitting of the Stern-Gerlach beams.  At time $t_2$, if the spin was
up, no action is taken by the observer.  If the spin was down, it
is either left
alone or flipped with some probability.  At time $t_3$, if the spin was
up at time $t_1$, the $z$-component of the spin is again determined.  In
Weinberg's non-linear quantum mechanics, the probability of the
measurement of the spin at time $t_3$, in the branch where
the spin was up at time $t_1$, depends on whether the observer did  or
did not flip the spin in the {\it alternative branch} where the spin was down
at time $t_1$ and the measurement at $t_3$ does not occur. That is the
``Everett phone''. In the language of the quantum mechanics of closed
systems this is simply an inconsistent set of histories.

These histories of the closed system spin and observer are
represented by a sequence of three  branch dependent sets of
projections at the times $t_1, t_2$, and $t_3$. They are  {\it branch}
{\it dependent} because, whether the alternatives $\{$flip, no flip$\}$ or the
trivial unit projection are used at time $t_2$ depends on the specific
alternatives $\{$up, down$\}$ at time $t_1$. Similarly the sets of projections
used at $t_3$ depend on the specific alternatives at time $t_1$.  Let
$\{P_\uparrow (t), P_\downarrow (t)\}$ be the projections representing
whether the spin is up or down at time $t$.  Let $\{P_f(t_2)$, $(P_{\bar f}
(t_2)\}$ represent the alternatives that the spin was flipped or not
flipped.  The four histories in the set described by Polchinski would be
represented by the chains
\begin{eqnarray}
P_\uparrow (t_3) I (t_2) P_\uparrow (t_1)\quad & ,\quad  I(t_3) P_f (t_2)
P_\downarrow (t_1)\ ,
\label{seveneight}\\
P_\downarrow (t_3) I (t_2) P_\uparrow (t_1)\quad & , \quad I (t_3) P_{\bar f}
(t_2) P_\downarrow (t_1)\ ,
\nonumber
\end{eqnarray}
where trivial unit projections have been included
 for clarity and vanishing
chains have been omitted.

Branch dependence is not an obstacle to defining the decoherence of a
set of histories \cite{Omnsum,GH90a,GH93a}. If the above set
decohered, the probabilities of the individual histories would be given
by
\begin{equation}
p(\alpha) = N\, Tr\left(XC_\alpha \rho\, C^\dagger_\alpha X^\dagger\right)
\label{sevennine}
\end{equation}
where $C_\alpha$, $\alpha= 1, 2, 3, 4$, is one of the four chains in
(\ref{seveneight}). The probabilities of histories in which
the spin is up or down at $t_3$ are independent of whether the spin was
flipped or not flipped at $t_2$ simply because the corresponding chains
contain neither the projection $P_f$ nor $P_{\bar f}$.

The above is a specific example of a general situation.  Consider a
decoherent set of branch dependent histories $\{\alpha\}$.  Partition
this set of histories into the class consisting of a single history
$\alpha$ and the class $\neg \alpha$ consisting of all other histories.
That partition is a coarse graining of the set $\{\alpha\}$ and so
is also decoherent. In the coarser-grained set,
the probability of $\alpha$ remains $p(\alpha)$.
The probability of $\neg\alpha$ is
\begin{equation}
p\bigl(\neg\alpha\bigr) = \sum_{\beta\not=\alpha} p (\beta)\ .
\label{seventen}
\end{equation}
Thus both $p(\alpha)$ and $p(\neg\alpha)$ are manifestly independent of
alternatives in the other branches. There are no ``Everett phones'' in
generalized quantum mechanics.  Decoherence guarantees the independence
of individual branches.

\section{conclusion}
\label{sec:conclusion}

The familiar quantum mechanics of unitarily evolving states on spacelike
surfaces depends centrally on the existence of a fixed background
spacetime geometry with a well-defined causal structure that is foliable
by spacelike surfaces on which the states can be defined.  When
spacetime geometry is not fixed, as in quantum gravity, or when it is
fixed but not foliable by spacelike surfaces, some modification of
familiar quantum theory seems inevitable.  Generalized quantum theory
provides a broad framework for constructing extensions of familiar
quantum theory that can apply when spacetime is not fixed \cite{Harpp}
 or when it is
 fixed but
not foliable by spacelike surfaces.  Such theories are unlikely
to permit
a notion of a unitarily evolving state on a spacelike surface or possess
familiar notions of causality.

In this paper we have discussed a
generalized sum-over-histories quantum mechanics for matter fields in
background spacetimes with non-chronal regions.  The geometry is fixed
and given once and for all time.  The matter fields do not modify it.
Alternatives are defined four-dimensionally as partitions of spacetime
field configurations --- a notion general enough to describe
alternatives in the non-chronal regions which are not foliable by
spacelike surfaces.  Transition amplitudes between alternatives on
spacelike surfaces outside the non-chronal regions are defined by sums
of $\exp[i({\rm action})]$ over intermediate field configurations.  The
non-unitarity of such transition amplitudes
can be incorporated into generalized quantum
theory through an appropriately defined notion of decoherence.  All
probability sum rules are satisfied for decoherent alternatives because
decoherence implies them.

This generalized quantum theory of fields in spacetimes with non-chronal
regions does not display a number of familiar features of quantum theory
in flat background spacetime.  Most importantly the theory cannot be
reformulated in terms of states on spacelike surfaces.  That is not
surprising since the spacetime itself does not possess a foliating
family of spacelike surfaces.  Lost with the notion of state is the
familiar idea of causality in the sense that the entire four-dimensional
spacetime geometry past, present, and future must be known to establish
the decoherence and predict the probabilities of alternatives in the
present.

Fundamentally spacetime geometry is not fixed but variable quantum
mechanically.  Quantum fluctuations in spacetime geometry are central to
a discussion of non-chronal regions because it is only through the
intervention of quantum gravity that spacetimes with non-chronal regions
could ever evolve \cite{Kli91,WY91}.  The present generalized quantum
mechanics of matter fields in a fixed background spacetime is thus only
a model or an approximation to a more general quantum mechanics
including the gravitational field.  It serves to illustrate, however,
how much of the structure of familiar quantum mechanics is tied to
assumptions concerning the character of spacetime geometry and what
departures from this structure we may expect in a generalized quantum
mechanics of geometry as well as matter fields.

\acknowledgments

The author is grateful to many scientists for conversations and
correspondence on these
issues and most especially to D.~Deutsch,
J.~Friedman, T.~Jacobson, J.~Preskill,
J.~Simon, K.~Thorne, S.~Trevedi, and U.~Yurtserver.  This work was
supported in part by the NSF under grant PHY90-08502.

\appendix
\section*{}
\label{sec:appendix}

In this Appendix we shall provide derivation and discussion of a few
results concerning the entropy functional used in Section VI.  We
consider operators $A$ that can be represented as
\begin{equation}
A= B\rho_1 B^{-1}
\label{aone}
\end{equation}
for some density matrix $\rho_1$.  We then define
\begin{equation}
\log A = B \log \rho_1 B^{-1}
\label{atwo}
\end{equation}
where $\log \rho_1$ is the Hermitian matrix that is diagonal in the basis
in which $\rho_1$ is diagonal with diagonal elements $\log p_i$ if
$p_i$ are the eigenvalues of $\rho$.  We may then construct the
entropy functional of $A$ and note that
\begin{eqnarray}
{\cal S} (A) = -Tr\bigl(A \log A\bigr)& = -Tr\left(B \rho_1 \log \rho_1
B^{-1}\right)
\nonumber\\
                                     & = -Tr\left(\rho_1 \log
\rho_1\right)\ . ~~~~~
\label{athree}
\end{eqnarray}
The inequality (\ref{sixeighteen})
\begin{equation}
-Tr\left(A \log \rho_2\right) \geq -Tr \bigl(A \log A\bigr)\ ,
\label{afour}
\end{equation}
where $\rho_2$ is any density matrix such that the left-hand side is
real and positive, may be proved by extremizing the functional $-Tr(A
\log C)$ over all operators $C$ with $Tr(C)=1$ (and thus over density
matrices in particular). Using a Lagrange multiplier $\mu$ to enforce
the constraint $Tr(C)=1$, the condition for an extremum is
\begin{equation}
\delta_C \Bigl[-Tr\bigl(A \log C\bigr) - \mu Tr (C) \Bigr] = 0\ .
\label{afive}
\end{equation}
This yields
\begin{equation}
AC^{-1} = \mu\ .
\label{asix}
\end{equation}
Taking $Tr(C)=1$ into account, this can only be satisfied by
\begin{equation}
C=A\ .
\label{aseven}
\end{equation}
since $Tr(A)=1$ from (\ref{aone}). Since $C=A$ is the unique extremum of
$-Tr(A \log C)$ among all operators with unit trace, it will also be an
extremum among operators $C$ such that $-Tr(A \log C)$ is real and
positive because, from (\ref{athree}), $-Tr(A \log A)$ is real and
positive.  To complete the demonstration of the inequality (\ref{afour})
it remains only to show that the extremum is a minimum.  However, it is
straightforward to exhibit density matrices $\rho_2$ for which $-Tr(A
\log \rho_2)$ is greater than $-Tr(A \log A)$ by using the basis
$|i\rangle$ in
which $\rho_2$ is diagonal with eigenvalues $\lambda_1$ to write
\begin{equation}
-Tr\left(A \log \rho_2\right) = \sum_i \left(-\log \lambda_i\right)
\bigl\langle i|A|i\bigr\rangle\ .
\label{aeight}
\end{equation}
If the dimension of the Hilbert space is a finite number $N$, choose
$\lambda_i = 1/N$.  Then, since $Tr(A)=1$ the left-hand side of
(\ref{aeight}) is just $\log N$.  This is the maximum value of ${\cal
S}(\rho)$ on density matrices and therefore greater than $-Tr(A \log
A)$ which, according to (\ref{athree}), is ${\cal S} (\rho_1)$.

The reader worried about the assumption of a finite dimensional Hilbert
space in the last argument can consider the following slightly more
special argument:  If there is a basis in which $\langle i|A|i\rangle$
is real and positive for a single basis vector $|i\rangle$, then we can
choose $\lambda_i=0$ for that vector to obtain a positive, infinite value
of $-Tr(A \log \rho_2)$ from (\ref{aeight}). The operator $A$ given by
(\ref{sixtwentyone}) certainly has such a basis, for example the basis
in which $\rho$ is diagonal. Then
\begin{equation}
\bigl\langle i|A| i \bigr\rangle = N p_i \bigl\langle i
|XX^\dagger
| i\bigr\rangle
\label{anine}
\end{equation}
where $p_i$ are the eigenvalues of $\rho$.  The right-hand side of
(\ref{anine}) is clearly real and positive.

Thus, the extremum $C=A$ is a minimum and the bound (\ref{afour}) is
established.

\begin{figure}
\caption{A compact non-chronal region of spacetime $NC$ with spacelike
surfaces $\sigma^\prime$ and $\sigma^{\prime\prime}$ before and after.
Alternatives may be defined on these spacelike surfaces, but the
transition matrix between them defined by a sum over intermediate field
configurations is not necessarily unitary if the field is interacting.}
\label{one}
\end{figure}

\begin{figure}
\caption{A local piece of a spacelike surface $R$ that is spacelike
separated from a non-chronal region $NC$. $R$ may be regarded either
 as lying
on a spacelike surface $\sigma^\prime$ before $NC$ or as lying on a
spacelike surface $\sigma^{\prime\prime}$ after $NC$.  If quantum
mechanics is to be consistently formulated in terms of states on
spacelike surfaces, then a prescription must be given for whether to
compute the probabilities of alternatives confined to $R$ with
$\sigma^\prime$ or $\sigma^{\prime\prime}$ if the evolution through $NC$
is not unitary for the results are not the same.}
\label{two}
\end{figure}

\begin{figure}
\caption{A spacetime with a single non-chronal region $NC$.  Before $NC$
there is an initial region that can be foliated by spacelike surfaces
some of which are illustrated.  Afterwards there is a similar final
region.
The text describes a generalized quantum mechanics for computing the
probabilities of decoherent histories of alternatives defined on these
surfaces even when the evolution through $NC$ is non-unitary.}
\label{three}
\end{figure}

\end{document}